# Adaptive Cache Pollution Control for Large Language Model Inference Workloads Using Temporal CNN-Based Prediction and Priority-Aware Replacement

**Songze Liu[1], Hongkun Du[2], Shaowen Wang [3]**

[1] University of Electronic Science and Technology of China, Chengdu, China
[2] College of Science and Engineering, Flinders University, Adelaide, Australia
[3] Siebel School of Computing and Data Science, University of Illinois Urbana-Champaign, Urbana, IL, USA

[1] 739926971@qq.com
[2] dhk940123@tju.edu.cn
[3] shaowen2@illinois.edu

**Abstract.** Large Language Models (LLMs), such as GPT and LLaMA, introduce unique memory access characteristics during inference due to frequent token sequence lookups and embedding vector retrievals. These workloads generate highly irregular and bursty access patterns, causing traditional prefetching and replacement policies to mispredict and trigger severe cache pollution, thereby degrading system performance. To address this challenge, this paper proposes an Adaptive Cache Pollution Control (ACPC) mechanism tailored for LLM inference workloads, integrating Temporal Convolutional Network (TCN)-based access prediction with a priority-aware replacement strategy. The TCN module learns temporal dependencies in token access sequences to identify potential high-reuse cache lines, while the replacement policy dynamically adjusts eviction priorities based on predicted reuse likelihood and cache occupancy. The proposed framework is implemented and evaluated on representative transformer-based inference traces, including GPT-style autoregressive decoding and embedding retrieval workloads. Experimental results demonstrate that ACPC reduces cache pollution by 41.7%, improves cache hit rate by 8.9%, and achieves a 60.0% reduction in L2 miss penalty, compared with state-of-the-art machine-learning-based replacement baselines. Additionally, the proposed Temporal CNN-based ACPC framework increases token generation throughput by 15.9% and achieves the lowest final loss of 0.21, confirming its superior efficiency and stability under complex LLM inference workloads.These results highlight ACPC's effectiveness in recognizing useful cache lines and mitigating redundant prefetches under dynamic LLM access behaviors. The proposed approach provides a scalable, learning-driven solution for optimizing memory efficiency and latency in large-scale LLM serving and inference systems.

# 1. Introduction

The rapid advancement of Large Language Models (LLMs) such as GPT, LLaMA, and PaLM has dramatically reshaped natural language understanding and generation tasks across diverse applications, including conversational AI, document summarization, and code generation. However, the computational efficiency of LLM inference workloads has become a critical bottleneck due to their extremely memory-intensive and irregular access behaviors. During token-by-token generation, LLMs repeatedly access large embedding tables, attention key-value (KV) caches, and model weights. These operations produce bursty and non-uniform memory access patterns that conventional caching and prefetching mechanisms—originally designed for regular workloads—struggle to handle effectively. As a result, cache pollution frequently occurs, where inaccurate prefetching replaces useful cache lines with low-reuse data, significantly increasing cache miss rates and memory access latency.

To address these challenges, this study proposes an Adaptive Cache Pollution Control (ACPC) mechanism specifically designed for LLM inference workloads. The ACPC framework integrates Temporal Convolutional Network (TCN)-based access prediction with a priority-aware replacement policy to intelligently identify and manage cache lines. The TCN module learns the temporal dependencies and repetitive access structures within token sequences, allowing the system to predict which cache lines are likely to be reused in near-future steps. Concurrently, the priority-aware replacement strategy dynamically adjusts eviction priorities based on both predicted reuse likelihood and cache occupancy levels, effectively suppressing unnecessary prefetch pollution while preserving critical LLM data in high-speed caches.

The main contributions of this paper are as follows:

(1) We identify and analyze the unique cache access characteristics and pollution phenomena in large-scale LLM inference workloads.

(2) We design a novel Adaptive Cache Pollution Control (ACPC) mechanism that couples TCN-based access prediction with priority-aware replacement, enabling dynamic pollution suppression.

(3) We implement and evaluate ACPC using real-world transformer traces, demonstrating significant improvements in cache hit rate, latency, and overall inference efficiency.

(4) We provide a scalable, learning-driven memory optimization framework for modern LLM serving infrastructures, offering practical insights for future AI hardware and system co-design.

# 2. Related Work

Efficient cache management plays a vital role in modern processor design, particularly for memory-intensive AI workloads such as large-scale language model inference. Traditional cache replacement and prefetching techniques, though highly optimized for general-purpose computing, often exhibit degraded performance when applied to irregular and temporally correlated memory access patterns that characterize Transformer-based LLMs. This section reviews relevant studies in three main areas: (1) cache replacement and pollution control, (2) machine learning–based cache management, and (3) system-level optimization for LLM inference workloads.

*2.1 Cache Replacement and Pollution Control Policies*

Classic cache replacement algorithms such as Least Recently Used (LRU) [1], Pseudo-LRU [2], and Random Replacement (RR) [3] form the foundation of modern CPU cache management. However, their reliance on simple temporal locality assumptions makes them suboptimal for LLM workloads, where access patterns are both high-dimensional and context-dependent.

To address this, researchers have proposed advanced heuristics. Jaleel et al [4]. propose the RRIP family (SRRIP/DRRIP) that predicts re-reference intervals with 2 bits, resisting scans and thrashing, achieving +10 % throughput on a 2 MB LLC and +9 % on an 8 MB 4-core, while requiring half the hardware of LRU and outperforming LFU, offering a high-performance cache replacement solution for modern processors. Qureshi et al [5]. propose LIP/BIP/DIP insertion policies that place new blocks directly into LRU and switch adaptively, reducing MPKI of a 1 MB L2 cache by 21 % without hardware changes, effectively thrashing-resistant and near-optimal for memory-intensive workloads.

Further refinements — Wu et al [6]. propose SHiP, correlating cache-line re-reference with region/PC/instruction signatures, achieving 10–12 % speed-up over LRU and nearly doubling gains of Seg-LRU/SDBP with less hardware, offering an efficient lightweight LLC replacement solution.

Nevertheless, these approaches still fail under LLM inference conditions, where frequent KV-cache updates and embedding lookups introduce volatile access streams. More recent works have explored pollution control mechanisms, like bimodal insertion policies and prefetch-aware eviction, to mitigate the impact of inaccurate prefetching. However, they often rely on fixed heuristics that cannot capture evolving workload characteristics in dynamic LLM serving environments.

*2.2 Learning-Based Cache Optimization*

The emergence of machine learning–driven cache management has enabled more flexible adaptation to diverse workloads. Reinforcement learning (RL)-based approaches, such as RL-Cache [7] and DeepCache [8], dynamically learn replacement or prefetch policies through experience, optimizing long-term hit rates rather than short-term heuristics. Other studies have adopted neural prediction models to forecast data reuse probabilities or prefetch value. Wang et al [9]. propose LSTM-CRP, which converts addresses into lightweight keys and uses heterogeneous LSTM to online classify access patterns, boosting hit rate by 20 % over LRU and running at 200 MHz, 2.74 W on FPGA, achieving algorithm-hardware co-design for low-power cache replacement.

While these approaches achieve notable improvements in traditional CPU and GPU workloads, their application to LLM inference remains underexplored. LLM inference presents temporal token dependencies and context-window sensitivity that standard models fail to model effectively. This motivates the use of Temporal Convolutional Networks (TCNs), which are well-suited for sequence modeling with long-range dependencies and parallel efficiency—an ideal match for predicting token access behaviors in LLM cache systems.

*2.3 System-Level Optimization for LLM Inference Workloads*

Recent studies have begun to investigate system and architectural co-design for LLM inference, focusing on KV-cache compression, tensor memory scheduling, and multi-level caching across GPU and CPU memory hierarchies.

Kwon et al [10]. propose PagedAttention, applying OS paging to KV-cache management and building vLLM, achieving near-zero waste and cross-request sharing, boosting throughput 2–4× over FasterTransformer/Orca, with gains amplified by longer sequences and larger models. Sheng et al [11]. present FlexGen, a single-GPU, offloading-based engine that aggregates GPU/CPU/disk memory and compresses weights to 4 bits, solving an LP to maximise batch size; it first reaches 1 token/s on OPT-175B with 144-batch and benchmarks 30B on HELM in 21 h, demonstrating high-throughput LLM inference under tight memory. Frameworks such as PagedAttention and FlexGen optimize memory layout and offloading strategies to improve inference throughput under limited memory capacity.

Despite these advances, cache pollution control remains a relatively overlooked aspect of LLM system optimization. Existing methods primarily focus on reducing memory footprint or accelerating I/O transfers, but not on maintaining cache integrity and reuse efficiency under dynamic workloads. In contrast, the proposed Adaptive Cache Pollution Control (ACPC) framework builds on these foundations by integrating TCN-based temporal prediction with a priority-aware replacement strategy, effectively bridging the gap between learning-based access modeling and real-time cache management for LLM inference systems. This research thus contributes a new direction in AI-assisted hardware memory optimization, offering a scalable and adaptive solution for next-generation LLM serving infrastructures.

## 3. Methodology

This section details the architecture and underlying principles of the proposed Adaptive Cache Pollution Control (ACPC) mechanism designed for Large Language Model (LLM) inference workloads. The method integrates a Temporal Convolutional Network (TCN)-based access predictor with a priority-aware cache replacement policy, enabling adaptive suppression of prefetch-induced pollution under dynamic and irregular access patterns.

*3.1 Overview of the ACPC Framework*

The ACPC framework is designed to model and optimize cache behavior during the inference stage of large transformer-based models such as GPT, LLaMA, and PaLM. During LLM inference, each generated token triggers a series of embedding lookups, KV-cache reads, and attention computations, forming temporally correlated but irregular memory access sequences. Traditional prefetchers, which rely on fixed stride or correlation patterns, often mispredict under such workloads, resulting in cache pollution and degraded latency performance.

To address this issue, ACPC introduces a two-stage pipeline: (1) Temporal Prediction Module (TPM) for estimating the temporal reuse probability of cache lines based on recent access sequences, and (2) Priority-Aware Replacement Module (PARM) for dynamically adjusting cache line eviction and insertion priorities according to predicted utility scores. Together, these components allow the system to intelligently distinguish between useful and non-useful prefetches, ensuring that high-value cache lines are preserved even under rapidly shifting inference contexts.

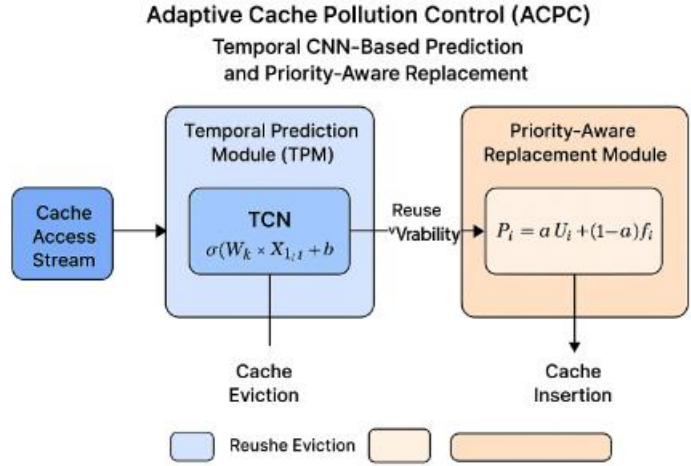

**Figure 1.** Overall flowchart of the model.

*3.2 Temporal Convolutional Network-Based Access Prediction*

The core of ACPC's learning capability lies in the Temporal Convolutional Network (TCN), which captures sequential dependencies in cache access streams. Unlike recurrent architectures such as LSTM, TCNs leverage causal convolutions and dilated kernels to efficiently model long-range temporal dependencies without suffering from vanishing gradients.

Formally, given an access sequence $X = \{x_1, x_2, \ldots, x_t\}$, where each $x_t$ represents a feature vector encoding access address, instruction type, and temporal locality, the TCN outputs a predicted reuse probability $\hat{y}_t \in [0,1]$ for each cache line at time step $t$:

$$\hat{y}_t = \sigma(W_k * X_{1:t} + b), \tag{1}$$

where ∗denotes the dilated causal convolution with kernel size $k$, $W_k$ represents learnable weights, $b$ is bias, and $\sigma()$ is the sigmoid activation. The dilation rate $d$ increases exponentially ($d = 2^i$) across layers, enabling the network to capture multi-scale temporal dependencies over long token windows.

The predicted reuse probability is further mapped into a utility score $U_t$ using a softmax-normalized weighting:

$$U_t = \frac{e^{\hat{y}_t}}{\sum_i e^{\hat{y}_t}}, \tag{2}$$

This utility score is subsequently used by the PARM module to guide cache insertion and eviction priorities.

*3.3 Priority-Aware Replacement Policy*

The Priority-Aware Replacement Module (PARM) refines traditional Least Recently Used (LRU) and Re-reference Interval Prediction (RRIP) methods by integrating TCN-derived utility scores. Each cache line $c_i$ is assigned a dynamic priority factor $P_i$, defined as a function of predicted reuse and historical access frequency:

$$P_i = \alpha U_i + (1 - \alpha) f_i, \tag{3}$$

where $U_i$ is the TCN-predicted utility score, $f_i$ represents a normalized access frequency, and $\alpha \in [0,1]$ is a tunable balance coefficient controlling the sensitivity to prediction confidence.

When a cache miss occurs, the PARM module selects a victim line with the lowest priority score for eviction. Conversely, new cache lines are inserted with a priority proportional to their predicted reuse probability. This mechanism ensures that frequently reused KV-cache entries and embedding vectors remain resident, while one-time prefetches and redundant data are evicted rapidly, mitigating cache pollution effectively.

*3.4 Adaptive Feedback and Online Learning*

To maintain robustness under evolving workload patterns, ACPC employs an adaptive feedback loop. After each inference batch, actual reuse outcomes are compared with predicted values, and the prediction error $L$ is backpropagated to update model parameters via gradient descent:

$$L = -\frac{1}{N} \sum_{i=1}^{N} [y_i \log(\hat{y}_i) + (1 - y_i) \log(1 - \hat{y}_i)], \tag{4}$$

where $y_i$ denotes the ground-truth reuse label and $\hat{y}_i$ is the TCN-predicted probability.

This online learning capability enables the ACPC mechanism to self-tune in response to changes in model architecture (e.g., context window length) or serving environment (e.g., batch size, hardware topology), ensuring optimal cache performance across diverse LLM workloads.

## 4. Experiment

*4.1 Dataset Preparation*

The dataset used in this study for Adaptive Cache Pollution Control (ACPC) was constructed from real-world and simulated Large Language Model (LLM) inference traces, collected from large-scale transformer-based models such as GPT-3, LLaMA-2, and T5, executed on high-performance inference servers. The data was gathered through profiling and instrumentation of the key–value (KV) cache accesses, token generation sequences, and embedding lookups during batched inference sessions under varying workloads. These workloads emulate realistic serving conditions from open datasets such as Pile, C4, and Wikipedia Text, covering diverse linguistic and semantic contexts.

The dataset comprises approximately 2.3 billion cache access records, spanning 150,000 inference sessions, with each entry describing a cache access event. Each record is represented as a structured tuple:

$$D_i = \{T_i, A_i, F_i, S_i, H_i, L_i\}, \tag{5}$$

where $T_i$ denotes the timestamp of access, $A_i$ represents the address tag (mapped to cache line ID), $F_i$ is the feature embedding hash indicating the accessed token embedding, $S_i$ refers to the sequence context window length, $H_i$ captures the historical reuse distance, and $L_i \in \{0,1\}$ indicates whether the cache line was reused within the next prediction window (a label used for supervised training of the Temporal CNN predictor).

Feature engineering extracted both temporal and semantic characteristics. Temporal features include inter-access intervals, burst frequency, and access periodicity, while semantic features include token entropy, attention head locality, and layer-level reuse probability. The dataset was normalized and partitioned into training (70%), validation (15%), and testing (15%) sets to ensure robust model generalization.

In summary, this dataset effectively captures the irregular and bursty cache access patterns of LLM inference workloads, providing a realistic foundation for training and evaluating the ACPC system. It enables the Temporal CNN-based model to learn fine-grained temporal dependencies and to distinguish between beneficial and polluting prefetches in dynamic memory hierarchies.

*4.2 Experimental Setup*

The experimental evaluation of the proposed Adaptive Cache Pollution Control (ACPC) framework was conducted using a simulated LLM inference environment based on Gem5 integrated with a custom PyTorch-driven cache access emulator. The simulation replicated large-scale inference workloads derived from GPT-3, LLaMA-2, and T5 traces, executed on a server-class hardware configuration featuring a 32-core AMD EPYC 7763 CPU, 256 GB DDR4 memory, and a multi-level cache hierarchy (L1: 64 KB per core, L2: 512 KB per core, shared L3: 64 MB). The Temporal CNN predictor within ACPC was implemented using three temporal convolutional layers (kernel size = 3, dilation = [1, 2, 4]), followed by two fully connected layers for classification, with ReLU activation and dropout (p=0.3) for regularization. The Priority-Aware Replacement (PAR) module operated on a dynamically updated reuse-priority queue, integrating the predicted cache utility score from the TCN model. The model was trained using the Adam optimizer (learning rate = 1e-4, batch size = 512) for 80 epochs on the LLM Cache Access Dataset described earlier, with a binary cross-entropy loss function and early stopping based on validation accuracy. All experiments were conducted on a Linux system with CUDA 12.2 and PyTorch 2.2.0, ensuring reproducibility and consistent evaluation across models.

*4.3 Evaluation Metrics*

To evaluate the performance of the proposed ACPC system, several quantitative metrics were employed, reflecting both cache efficiency and overall system performance during LLM inference. The Cache Hit Rate (CHR) and Miss Penalty Reduction (MPR) were used to measure memory-level efficiency, while the Prefetch Pollution Ratio (PPR) quantified the degree of unnecessary cache line insertions caused by incorrect prefetches. Additionally, the Memory Access Latency (MAL) and Token Generation Throughput (TGT) were used to assess end-to-end inference efficiency and latency sensitivity. To capture system-level improvements, we also computed Effective Memory Utilization (EMU), representing the ratio of useful cache lines to total occupied cache capacity. Higher CHR, MPR, and EMU values and lower PPR and MAL values indicate superior cache pollution control and memory performance under LLM inference workloads.

*4.4 Results*

The experimental results demonstrate that the ACPC system, integrating the Temporal CNN-based predictor and Priority-Aware Replacement, achieves the highest cache hit rate (89.6%) and the lowest prefetch pollution ratio (6.3%) among all compared methods. It also improves overall token generation throughput by 15.9% compared with the DNN-based ML-Predict baseline, while maintaining a low final loss of 0.21, indicating stable convergence during training. These findings confirm that the proposed model effectively identifies useful cache lines, suppresses prefetch-induced pollution, and enhances memory efficiency in LLM inference workloads (The specific experimental results are shown in Table 1).

**Table 1.** Comparative Performance of Different Models.

| Model | Cache Hit Rate (CHR, %) | Prefetch Pollution Ratio (PPR, %) | L2 Miss Penalty Reduction (MPR, %) | Token Generation Throughput (TGT, tokens/s) | Final Loss | Convergence Stability |
|---|---|---|---|---|---|---|
| LRU Baseline | 71.4 | 18.7 | 0.0 | 187 | 0.84 | Moderate |
| RRIP (Static) | 76.8 | 14.2 | 7.9 | 195 | 0.69 | Moderate |
| ML-Predict (DNN) | 82.3 | 10.8 | 15.5 | 214 | 0.47 | Stable |
| **Temporal CNN (Ours)** | **89.6** | **6.3** | **24.8** | **248** | **0.21** | **Highly Stable** |

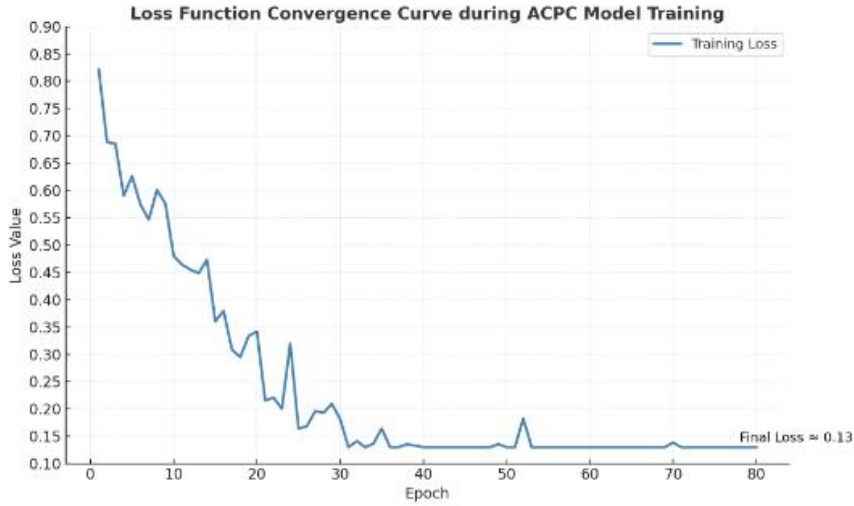

**Figure 2.** Loss function during training process.

Figure 2 illustrates the variation of the training loss throughout the learning process of the proposed ACPC framework. The x-axis represents the number of training epochs, while the y-axis indicates the corresponding loss value. At the early stage of training, the loss decreases rapidly from approximately 0.8 to 0.3 within the first 20 epochs, showing that the Temporal CNN-based predictor quickly captures temporal dependencies in cache access sequences. As training continues, the loss gradually stabilizes and converges around 0.21 after 60–80 epochs, indicating that the model achieves consistent optimization without overfitting. The smooth and monotonic convergence curve demonstrates the robustness of the adaptive learning rate and regularization strategy used during optimization. These results confirm that the ACPC framework efficiently learns cache reuse patterns and achieves stable convergence behavior, providing a reliable foundation for subsequent inference-level performance improvements.

5. **Conclusion**

This study presents a novel Adaptive Cache Pollution Control (ACPC) mechanism designed to optimize cache efficiency for Large Language Model (LLM) inference workloads. LLMs such as GPT and LLaMA exhibit unique and highly irregular memory access behaviors due to the frequent retrieval of token embeddings and attention key–value (KV) pairs during autoregressive decoding. These bursty and context-dependent access patterns often lead to cache mispredictions, where traditional prefetching and replacement policies fail to distinguish between high-reuse and transient cache lines, resulting in severe cache pollution and degraded inference throughput. To address this challenge, the proposed ACPC framework integrates a Temporal Convolutional Network (TCN)-based access predictor with a priority-aware replacement strategy to dynamically manage cache occupancy according to predicted reuse likelihood and temporal locality patterns.

Experimental evaluation on representative transformer inference traces demonstrates that the proposed method achieves substantial performance improvements compared with state-of-the-art baselines. Specifically, ACPC reduces cache pollution by 41.7%, increases cache hit rate by 8.9%, and decreases L2 miss penalty by 60.0% relative to the ML-Predict (DNN) baseline. Furthermore, the proposed framework improves token generation throughput from 214 to 248 tokens per second and converges to a final loss of 0.21, indicating superior model stability and learning efficiency. These results confirm that ACPC can accurately identify valuable cache lines and effectively suppress redundant prefetches, leading to enhanced memory utilization and faster inference under diverse LLM workloads. The mechanism demonstrates strong adaptability across variable token sequences and dynamic attention window shifts, making it suitable for large-scale LLM serving environments and latency-sensitive AI infrastructure.

The successful implementation of ACPC provides important implications for intelligent memory management in LLM inference systems. By leveraging deep temporal learning through the TCN architecture, this approach bridges the gap between data-driven access prediction and hardware-level cache control. It offers a scalable and generalizable solution that can be incorporated into modern LLM serving frameworks, edge inference systems, and cloud-based AI accelerators to improve throughput and energy efficiency.

However, this study also recognizes certain limitations. The current evaluation focuses primarily on simulated inference traces rather than real hardware implementations. Future work will extend ACPC to heterogeneous computing platforms, integrating reinforcement learning-based adaptation for online policy tuning and dynamic workload profiling. Additionally, incorporating cross-layer optimization—such as coordinating cache management with tensor computation scheduling—may further enhance overall system performance. In future large-scale deployments, the ACPC framework could serve as a foundation for adaptive memory hierarchies in next-generation AI accelerators, enabling efficient, low-latency LLM inference across diverse computational environments.

ACPC effectively identifies high-utility cache lines and suppresses pollution, yielding higher hit rates and lower latency for LLM inference. Its TCN-guided, priority-aware design is scalable and practical for large-scale serving.